\documentclass[twocolumn,showpacs,preprintnumbers,amsmath,amssymb]{revtex4-1}
\usepackage{graphicx}
\usepackage{dcolumn}
\usepackage{bm}

\newcommand{\be}{\begin{eqnarray}}
\newcommand{\ee}{\end{eqnarray}}

\begin{document}


\title{Dynamics of quantized vortices in Bose-Einstein condensates with laser-induced spin-orbit coupling}

\author{Kenichi Kasamatsu}
\affiliation{
Department of Physics, Kinki University, Higashi-Osaka, 577-8502, Japan \\ and Institut f\"{u}r Theoretische Physik, Leibniz Universit\"{a}t Hannover, 30167 
Hannover, Germany
}

\date{\today}

\begin{abstract}
We study vortex dynamics in trapped two-component 
Bose-Einstein condensates with a laser-induced spin-orbit coupling using 
the numerical analysis of the Gross-Pitaevskii equation. 
The spin-orbit coupling leads to three distinct ground state phases, which depend on some 
experimentally controllable parameters. When a vortex is put in one or both of 
the two-component condensates, the vortex dynamics exhibits very different behaviors 
in each phase, which can be observed in experiments. 
These dynamical behaviors can be understood by clarifying the stable vortex structure 
realized in each phase. 
\end{abstract}

\pacs{03.75.Lm, 03.75.Mn, 67.85.Fg}
\maketitle

\section{Introduction} \label{intro}
Experimental realization of spin-orbit (SO) coupled Bose-Einstein condensates (BECs) 
in cold atoms have opened a new avenue of studying rich dynamical behaviors in new quantum 
fluids \cite{Lin11,Galitskirev,Goldmanrev,Zhairev}.  
The coupling of the momentum and spin degrees of freedom 
may yield new structure of topological defects. 
Quantized vortices in atomic BECs have been thoroughly studied 
for past decades \cite{Fett09}, but those in SO coupled BECs are a quite new topic. 
It has been shown that a coreless vortex referred to as a half-quantized vortex 
(HQV) can be the ground state for a small condensate with a 
symmetric two-dimensional (2D) SO coupling known as the Rashba-type \cite{Wu,Sinha,Hui}. 
Also, some studies considered the vortex structures in SO coupled BECs in 
a rotating trap \cite{Radi11,Xu,Zhou}. 
 
Experimentally, the SO coupling can be synthesized by the laser-induced Raman coupling 
between different hyperfine states of atoms, which realizes a 1D SO coupling 
corresponding to the combination of equal weight of the Rashba- and Dresselhaus-type 
coupling \cite{Lin11,Zhang,Wang,Cheuk}; 
a recent experiment has demonstrated creation of a 2D synthetic SO coupling for 
cold atoms \cite{Huang}.
The laser-induced SO coupling realized in the experiments provides rich physics 
of the BECs and much of these properties have been studied \cite{Zhairev,Ho,Li,Martone,Lu13,Li2,Ozawa,Hamner}.  
Recently, Fetter has considered the dynamics of a single vortex in trapped BECs 
with the laser-induced SO coupling, giving characteristic features of the dynamics based 
on the time-dependent Lagrangian formalism \cite{Fetter}. 
Although there remains an experimental difficulty to create a vortex in the SO coupled condensate 
by using a rotating trap as done before \cite{Radi11},
the vortex creation without rotation may be possible by
the rapid thermal quench of the atomic gas into the condensation regime \cite{Fetter,Weiler}. 
Real time dynamics of vortices was observed through the snapshots 
of the condensate density, where successive short microwave pulses were applied to 
transfer a small fraction of condensed atoms to the untrapped state \cite{Frei10}.  

In this paper, we study the vortex dynamics in laser-induced SO coupled BECs through the 
numerical simulations of the 2D Gross-Pitaevskii (GP) equation. 
There are three distinct phases for the BECs with the Raman-induced SO coupling, namely 
the stripe phase, the plane-wave phase, and the mixed phase \cite{Zhairev,Ho,Li}. 
These phases can be stabilized  
by changing the Rabi frequency that depends on the strength of the Raman laser beam. 
We study the vortex dynamics when a single vortex is put in one or both of the components.  
The vortex structure in each phase is strongly dependent on the 
energetic constraint caused by both the SO coupling and the Rabi coupling. 
As a result, vortex dynamics exhibits a quite different behavior in each phase. 

This paper is organized as follows. 
Section \ref{lagrangian} introduces the basic formulation of the SO coupled 
BECs and addresses briefly the ground state property of this system, 
which is necessary to understand the following discussion.
In Sec.~\ref{nume}, we show results of our numerical simulations 
of the vortex dynamics in the three phases of the SO coupled BEC. 
Section~\ref{concle} devoted to the discussion and conclusion. 

\section{BASICS}\label{lagrangian}
\subsection{Formulation for Raman-induced SO coupled BECs}
We consider a 2D system of two-component (psudospin-1/2) BECs 
with a laser-induced SO coupling \cite{Lin11} to study the vortex dynamics. 
The single-particle SO Hamiltonian has the $2\times 2$ matrix structure 
\begin{equation}
h_0 = \frac{\hbar^2}{2m} \left(-i\bm \nabla \sigma_0+ k_0 \hat{\mathbf{x}} \sigma_z \right)^2 
+ \frac{\hbar \delta}{2} \sigma_z+ \frac{\hbar\Omega}{2} \sigma_x + V_\mathrm{tr} \sigma_0. 
\label{singleho}
\end{equation}
Here, $m$ is the atomic mass, $\sigma_r$ for $r=x,y,z$ is one 
of the Pauli matrices and $\sigma_0$ is the unit matrix.  
The trapping potential is assumed to be a harmonic form 
$V_\mathrm{tr} = m \omega_{\perp}^2 r^2 / 2$. 
As discussed in Ref.~\cite{Lin11}, this spinor Hamiltonian has three 
parameters under experimental control: $k_0$ is the wavenumber of the 
Raman laser beams, $\Omega$ is the associated Rabi frequency related 
to the intensity of the laser beams, and $\delta$ is the detuning 
controlled by an external magnetic field. 
The kinetic energy term has a uniform synthetic gauge field 
$ -\hbar k_0 \hat{\mathbf{x}} \sigma_z$ proportional 
to the spin matrix $\sigma_z$. This term represents the 1D SO coupling 
along the $x$-direction whose magnitude can be controlled by $k_0$. 

The Lagrangian and the energy functional including the single-particle SO Hamiltonian 
of Eq.~(\ref{singleho}) is given by
\begin{align}
L &= \int d^2 r \frac{i \hbar}{2} \left( \Psi^{\dagger} \frac{\partial \Psi}{\partial t} 
- \frac{\partial \Psi^{\dagger}}{\partial t} \Psi \right) - E[\Psi], \label{lagGP} \\
E[\Psi] &= \int d^2r\left( \Psi^\dagger h_0 \Psi + 
\frac{g}{2} \sum_{j=1,2} |\Psi_j|^4 + g_{12} |\Psi_1|^2 |\Psi_2|^2 \right). \label{EGPSO}
\end{align} 
The order parameters are represented by the 2-component spinor $\Psi = (\Psi_1,\Psi_2)^{T}$.
Here, we set the same intracomponent coupling constant as $g$ for simplicity, 
and gives $g_{12} / g \equiv \gamma = 0.995$ according to the parameters of the $^{87}$Rb atoms \cite{Lin11}. 
This small difference between $g$ and $g_{12}$ is important to stabilize the 
stripe phase in the SO coupled BECs \cite{Li}. 
We scale the energy using the recoil energy $E_R = \hbar^2k_0^2/2m$ 
with the wave number $k_0$ of the Raman photon. 
Taking the length scale by $k_0^{-1}$, we have 
\begin{align}
\tilde{h}_0 = (-i \nabla \sigma_0 + \hat{\mathbf{x}} \sigma_z)^2 + \frac{\tilde{\delta}}{2} \sigma_z 
+ \frac{\tilde{\Omega}}{2} \sigma_x + \tilde{V}_{\mathrm{tr}} \sigma^0,
\end{align}
where the dimensionless quantities have tildes. 
The coefficient of the trap potential becomes $(a_{\mathrm{ho}} k_0)^{-4}$ with 
the harmonic oscillator length $a_{\mathrm{ho}} = \sqrt{\hbar/m\omega_{\perp}}$. 
Since $a_{\rm ho} $ is a few times larger than $2\pi/k_0$ in a usual experimental situation, 
the coefficient of the trap potential becomes very small; we use $(a_{\mathrm{ho}} k_0)^{-4} = 0.005$ 
in the following calculation.  
The normalization of the wave function is given by the total particle number in the 2D system 
$N = \int d^2r \Psi^{\dagger} \Psi =  \int d^2r (|\Psi_1|^2 + |\Psi_2|^2) = N_1 + N_2$. 
By replacing the wave function as $\Psi = \sqrt{N} k_0 \tilde{\psi}$, 
where $\int d^2r |\tilde{\psi}|^2 =1$, one can  
define the dimensionless 2D coupling strength as $\tilde{g}_{2d} = 2 m g N / \hbar^2$ \cite{tyyu}. 
The time-dependent GP equations derived from Eq.~(\ref{lagGP}) 
can be written as  
\begin{align}
i \frac{\partial \psi_1}{\partial t} = -(\nabla^2 + 2 i \partial_x) \psi_1 + \frac{\delta}{2} \psi_1 + \frac{\Omega}{2} \psi_2 + V_{\mathrm{tr}} \psi_1 \nonumber \\
+ g_{2d} (|\psi_1|^2 + \gamma |\psi_2|^2) \psi_1,   \label{tdgp1} \\
i \frac{\partial \psi_2}{\partial t} = -(\nabla^2 - 2 i \partial_x) \psi_2 - \frac{\delta}{2} \psi_2 + \frac{\Omega}{2} \psi_1 + V_{\mathrm{tr}} \psi_2 \nonumber \\
+ g_{2d} (|\psi_1|^2 + \gamma |\psi_2|^2) \psi_2,  \label{tdgp2}
\end{align}
where tildes are omitted in the notation. 
In the following, we fix $g_{2d} = 1000$ and $\delta = 0$ for simplicity. 
Thus, the free parameter in our study is only the Rabi frequency $\Omega$.

\subsection{Brief review of the ground state phases without a vortex}
\begin{figure}[ht]
\centering
\includegraphics[width=0.9\linewidth,bb=0 0 368 567]{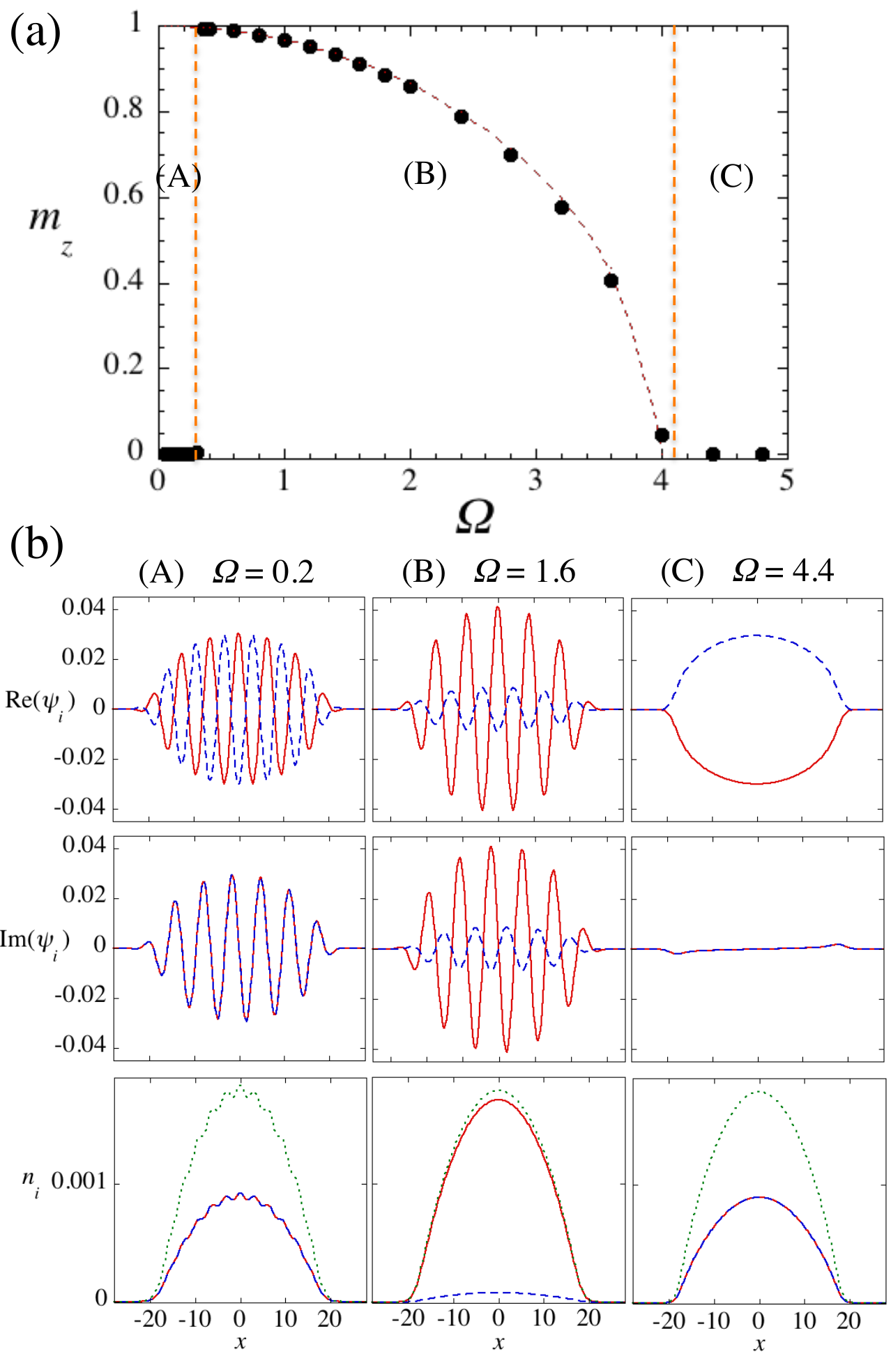} \\\
\caption{(Color online) The phases of the ground states of the BEC with the laser-induced SO coupling. 
(a) The population difference $m_z$ between two components as a function of $\Omega$. 
The phases (A), (B), and (C) correspond to the stripe, plane-wave, and mixed phases, respectively. 
A dotted curve in the region (B) represents Eq.~(\ref{planek1}). 
The result is consistent with that of the 1D GP equation reported in Ref.~\cite{Li}. 
(b) The cross section along $y=0$ of the condensate wave function in each phase: 
the real part (upper), the imaginary part (middle), and the density in each component (bottom). 
The solid and dashed curves correspond to the quantities for 
$\psi_1$- and $\psi_2$-component, respectively. The total density is also shown in the bottom by 
dotted curves. }
\label{population}
\end{figure}

The starting point in our study is to understand the ground state structure without vorticity. 
Figure \ref{population} summarizes the results obtained by solving Eqs.~(\ref{tdgp1}) and (\ref{tdgp2}) 
via imaginary time evolution. 
The ground state can be distinguished by the population difference (longitudinal spin polarization) 
$m_z=|N_1-N_2|/N$ with respect to $\Omega$ as, (A) the stripe phase, 
(B) the plane-wave (polarized) phase, and (C) the mixed (single-minimum) phase \cite{Lin11,Ho,Li}. 
These phases are determined by the properties of the minima in the single-particle 
spectrum of the hamiltonian Eq.~(\ref{singleho}). The stripe phase is a consequence 
in which the condensed particles are equally populated to two degenerate 
minima of finite linear momenta with an opposite sign. 
The plane-wave phase occurs when the particles condense to either of the two degenerate minima. 
With increasing $\Omega$, the two minima merge to a single minimum, which 
results in the mixed phase with a zero linear momentum. 

The condensate wave function of this system can be written as 
$\psi = \sqrt{n(\mathbf{r})} \eta$ with the spinor $\eta$ being approximated by the ansatz \cite{Li}
\begin{equation}
\eta = 
C_{+} \left( 
\begin{array}{c}
\sin \theta \\
-\cos \theta 
\end{array}
\right) e^{+i k_1 x}
+
C_{-} \left( 
\begin{array}{c}
\cos \theta \\
-\sin \theta 
\end{array}
\right) e^{-i k_1 x}, 
\end{equation}  
where $k_1$ represents the canonical momentum (in unit of $k_0$) where the Bose-Einstein 
condensation takes place and $C_{\pm}$ is the complex amplitude of the $\pm k_1$ component. 
The Rabi coupling favors the $\pi$ phase difference between two components, 
so that the $\psi_2$-component has a minus sign. 
The angle $\theta$ can be determined by $\cos (2 \theta) = k_1$, given by the 
energy minimization with respect to $k_1$ \cite{Li}. 
The total density $n(\mathbf{r})$ is approximately given by the Thomas-Fermi profile \cite{Abad}
\begin{equation}
n(r) = \frac{\mu - V_{\mathrm{tr}}}{g_{2d}(1+\gamma)/2-\Omega/(4|\psi_1||\psi_2|)}. \label{TFrabi1}
\end{equation}
Although in the stripe phase the total density $n(\mathbf{r})$ is periodically modulated slightly, 
Eq.~(\ref{TFrabi1}) is still good to represent the smoothed global density profile. 
For $\Omega \ll g_{2d}$ the usual Thoms-Fermi profile for a single-component system 
is reproduced as $n(r) = n(0) (1-r^2/R_{\perp}^2)$, 
where $n(0) = 2 \mu/g_{2d}(1+\gamma)$ is the condensate density at the origin ($r=0$) and 
$R_{\perp}=\sqrt{\mu} (k_0 a_{\mathrm{ho}})^2$ is the Thomas-Fermi radius. 
The chemical potential is determined by the normalization condition. 

The stripe phase is realized for small $\Omega$ $(\leq 0.35)$ in our parameter setting. 
The wave function can be described by $|C_{+}| = |C_{-}| = 1/\sqrt{2}$ and the wave number $k_1$ given by 
\begin{equation}
k_1 = \sqrt{1-\frac{\Omega^2}{16(1 + G_1)^2}}, 
\end{equation} 
where $G_1=\bar{n} g_{2d}(1+\gamma)/8$ with the mean density $\bar{n} = \int d^2 r n^2(r)$. 
For small $\Omega \ll 1$, one can approximate $k_1 \approx 1$ and thus $\theta = 0$ or $\pi$. 
We can eventually write the wave function as 
\begin{equation}
\psi \simeq \frac{n(r)}{\sqrt{2}} \left( 
\begin{array}{c}
e^{-i x} \\
- e^{i x} 
\end{array}
\right) . \label{stripesimple}
\end{equation}
Here, we take the relative phase between $C_{+}$ and $C_{-}$ to be zero. 
Figure~\ref{population}(b-A) shows that 
the wave function can be described by Eq.~(\ref{stripesimple}), where the 
real and imaginary parts are out-of-phase and in-phase between the two components, 
respectively, with respect to $x$. 

Next, we consider the plane-wave (polarized) phase, where the condensation takes place 
in a single plane-wave state with either $+k_{1}$ or $-k_{1}$; we choose 
$+k_1$ in the discussion here. Then, $C_+=1$ and $C_-=0$ and the wave function 
is given by 
\begin{equation}
\psi = n(r)  \left( 
\begin{array}{c}
\sin \theta \\
- \cos \theta
\end{array}
\right) e^{i k_1 x} ,
\label{planewavewav}
\end{equation}
and 
\begin{equation}
k_1 = \sqrt{1-\frac{\Omega^2}{16(1 - 2 G_2)^2}}  = m_z
\label{planek1}
\end{equation}
with $G_2=\bar{n} g_{2d}(1-\gamma)/8$.
In our case, $G_2$ is very small parameter and actually neglected. 
The signature of the plane-wave phase is the finite population difference, 
which actually follows the relation $m_z = k_1$ as shown in Fig.~\ref{population}(a).
The imaginary part of the wave function becomes out-of-phase [Fig.~\ref{population}(b-B)], 
which is also a clear difference from the stripe phase. 

The third phase is the mixed (single minimum) phase, where condensation takes place 
at zero momentum $k_1 = 0$. Then, the wave function does not exhibit 
sinusoidal oscilation as seen in Fig.~\ref{population}(b-C), so that the average 
spin polarization vanishes.  

\section{Numerical simulation of vortex dynamics}\label{nume}
Here, we numerically simulate the vortex dynamics in SO coupled BECs by solving 
the 2D GP equations (\ref{tdgp1}) and (\ref{tdgp2}) \cite{chu2}. 
We prepare a vortex at a certain position in the condensate in the following way. 
First, we calculate the ground state wave function without vortices by imaginary time 
evolution of Eqs.~(\ref{tdgp1}) and (\ref{tdgp2}). 
Next, we imprint a phase defect at a certain position 
$\mathbf{r}=\mathbf{r}_v$ by multiplying the phase $e^{i\phi}$ with the profile 
$\phi(\mathbf{r}-\mathbf{r}_v) = \arctan(\frac{y-y_v}{x-x_v})$ in one or both of the components. 
Using this wave function as the initial state, we again proceed short imaginary time evolution, 
which can make the amplitude of the wave function converge to the proper profile 
without changing the vortex position significantly. 
As a result, we can prepare the initial state of real time evolution \cite{chu1}. 
Here, we denote the winding number $q_1$ and $q_2$ of the vortex in the 
$\psi_1$- and $\psi_2$-component, respectively. 

\subsection{Stripe phase} \label{stripevortexdy}
\begin{figure}[ht]
\centering
\includegraphics[width=1.0\linewidth,bb=0 0 487 368]{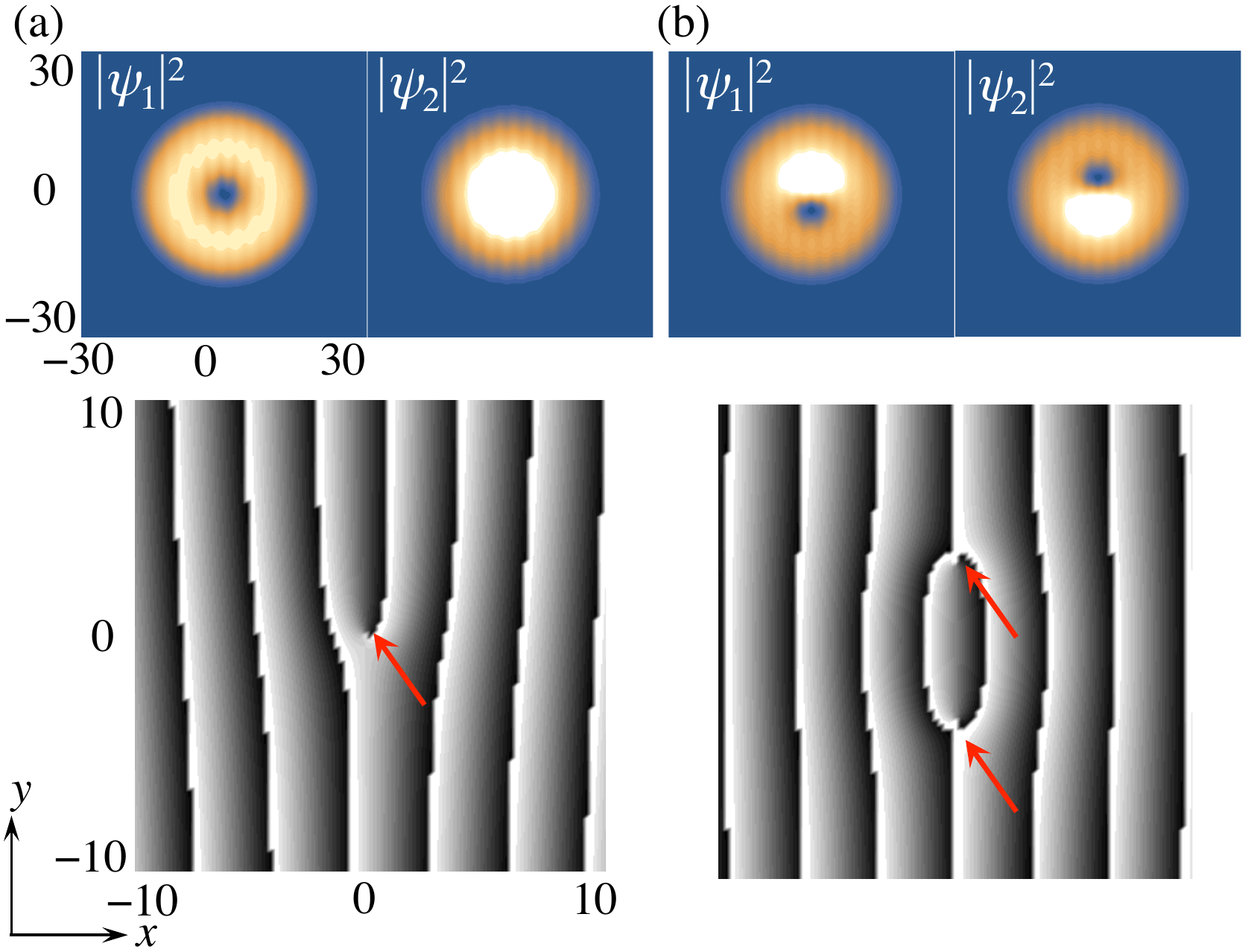} \\\
\caption{(Color online) Typical initial states used in the simulations to study the vortex dynamics 
in the stripe phase for $\Omega = 0.2$. The vortex winding number 
is (a) $(q_1,q_2) = (1,0)$ and (b) $(q_1,q_2) =(1,1)$. The upper and lower panels 
represent the density profile of each component and the relative phase $\theta_1 - \theta_2$. 
The phase value changes from $-\pi$ (black) to $\pi$ (white) continuously.  
The vortex is represented in the relative phase profile by arrows.}
\label{stripeinitial}
\end{figure}

First, we discuss the vortex dynamics in the stripe phase, 
where we choose $\Omega = 0.2$. In the simulation, we consider two situations 
shown in Fig.~\ref{stripeinitial} as the typical initial states, where the winding number of the vortex 
in each component corresponds to $(q_1,q_2) = (1,0)$ and $(1,1)$. 

From these initial states, we can find some important properties for laser-induced 
SO coupled BECs. 
Since the two-component are coupled by the uniform coherent Rabi coupling, one 
cannot make a vortex in one of the components without affecting the phase in the other component, 
because the relative phase between the two components likes to be uniform 
to decrease the Rabi coupling energy.  
In the stripe phase, however, the Rabi coupling energy gives a minor contribution, 
which enable us to make vortices freely in each component. 
This can be seen in the phase-dependent energy of the system. 
When we put one vortex in each phase by multiplying the phase $\phi_i=\phi(\mathbf{r} - \mathbf{r}_{vi})$ ($i=1,2$), 
the kinetic energy of the system is written as 
\begin{align}
E_{k} = \int d^2 r n(r) \biggl\{ |C_+|^2 \sin^2 \theta \left[ (v_{1x}+k_1+1)^2 + v_{1y}^2 \right] \nonumber \\
+ |C_-|^2 \cos^2 \theta \left[ (v_{1x}-k_1+1)^2 + v_{1y}^2  \right] \nonumber \\ 
+ |C_+|^2 \cos^2 \theta \left[ (v_{2x}+k_1-1)^2 + v_{2y}^2 \right] \nonumber \\
+ |C_-|^2 \sin^2 \theta \left[ (v_{2x}-k_1-1)^2 + v_{2y}^2  \right]
\biggr\} , \label{kinesocene}
\end{align}
where $\mathbf{v}_i = \nabla \phi_i = (v_{ix},v_{iy})$ is the velocity field induced by a vortex 
in the $i$-th comonent. On the other hand, 
the energy of the Rabi coupling is 
\begin{equation}
E_{\mathrm{Rabi}} = - \frac{\Omega}{2} \int d^2 r n(r) \sin2\theta \cos(\phi_1 -\phi_2).
\end{equation}
For the stripe phase, $|C_+| = |C_-| = 1/\sqrt{2}$ and $k_1 \approx 1$, i.e., $\theta \approx 0$. Then, 
$E_{\mathrm{Rabi}}$ is almost zero and does not affect the system even if the relative phase 
is strongly disturbed. Thus, vortices can be created without costing the 
Rabi coupling energy. There is in principle no problem for the initial vortices to have 
combinations of arbitrary values of $q_1$ and $q_2$. 
The kinetic energy becomes $E_{k} \approx \frac{1}{2}(v_1^2+v_2^2)$, which is 
an usual kinetic energy when vortices are put on usual two-component condensates. 
Therefore, vortex dynamics in the stripe phase can be approximately regarded as those in 
two-component BECs with a very weak coherence of the relative phase caused 
by a small Rabi coupling. 

The above argument can be interpreted from the relative phase profile 
shown in the bottom of Fig.~\ref{stripeinitial}. 
The stripe phase already has a gradient $-2x$ in the relative phase 
because of the relatively weaker Rabi coupling than the SO coupling. 
Therefore, it is possible to create a single defect like $(q_1,q_2) = (1,0)$ 
[Fig.~\ref{stripeinitial}(a)], where this vortex can be seen as 
a local dislocation in the relative phase profile as 
shown in Fig.~\ref{stripeinitial}(a); the $(q_1,q_2) = (1,0)$ vortex is not allowed 
for other two phases, because such a single defect must break the uniform relative 
phase globally. On the other hands, the $(q_1,q_2) = (1,1)$ vortex can 
be created with less damage of uniformity in the relative phase, as seen in Fig.~\ref{stripeinitial}(b). 
This is interpreted as the binding of two vortices by the sine-Gordon soliton 
of the relative phase \cite{Son} to form a vortex molecule in 
the coherently coupled BEC \cite{Kasamatsu,Kasamatsu2,Cipriani}. 
Although the circulations of the vortices are seen to be anti-parallel in the relative phase, 
this is a vortex pair with the same direction of circulations. 

\begin{figure}[ht]
\centering
\includegraphics[width=1.0\linewidth,bb=0 0 720 340]{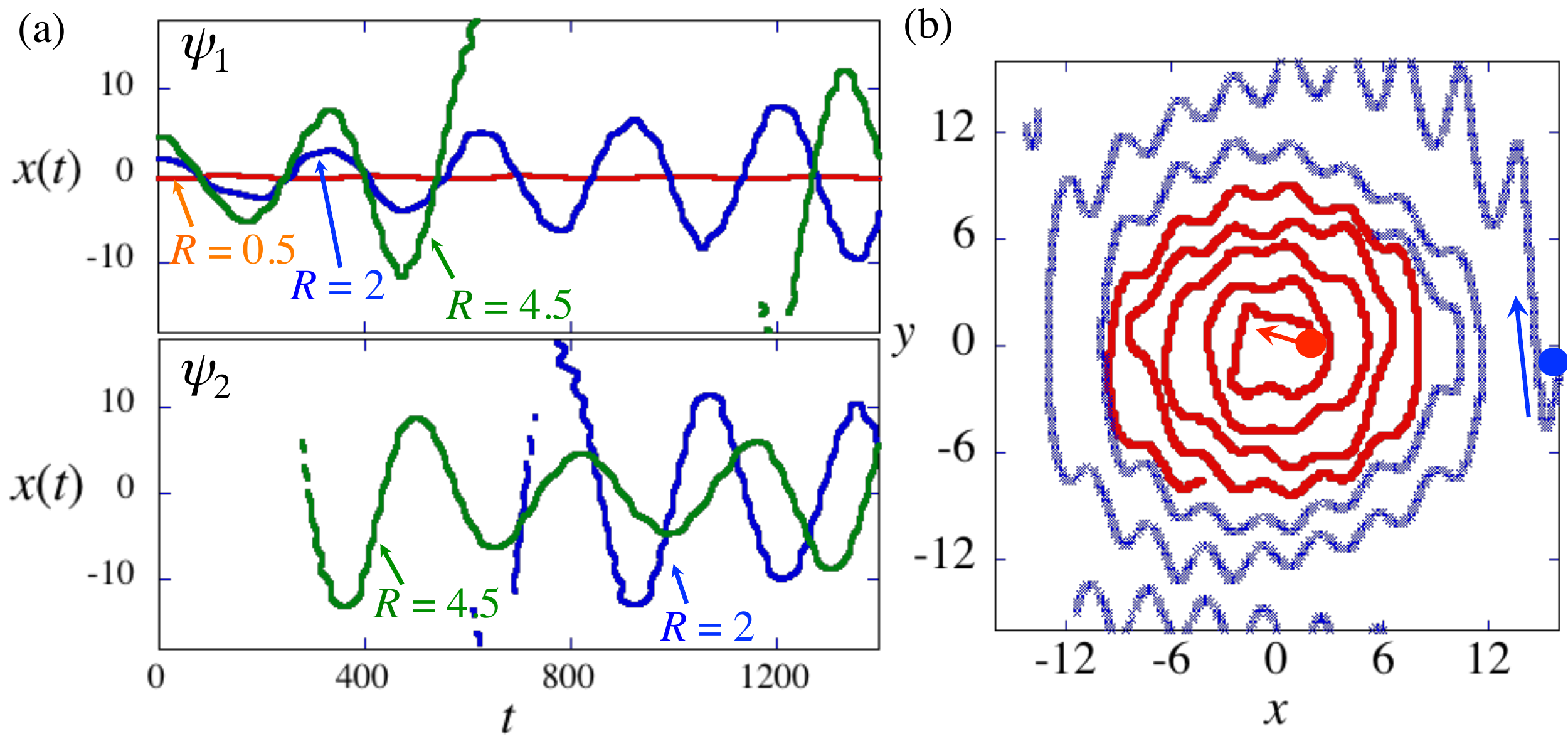} \\\
\caption{(Color online) The time development of the vortex state with the winding number 
$(q_1,q_2) = (1,0)$ and $\Omega=0.2$. 
(a) The time development of the $x$-coordinate of the vortex in 
the $\psi_1$-component (upper panel) and the $\psi_2$-component (lower panel) 
for the different initial displacements $R=0.5$, $2.0$ and $4.5$ from the center. 
(b) The vortex trajectories in the $\psi_1$-component 
(from center to outer: the red curve) and $\psi_2$-component 
(from outer to center: the blue curve) for the initial displacement $R=2.0$.}
\label{10vortexstripe}
\end{figure}
First, we show in Fig.~\ref{10vortexstripe} the numerical result of vortex dynamics for the $(q_1,q_2) = (1,0)$ vortex state, 
where we change the initial position $\mathbf{r}_{vi} = (R,0)$ of the vortex. 
The positive displacement $R$ from the trap center induces counterclockwise precession 
of a vortex around the center \cite{Fett09}. 
For a centered vortex or a slightly off-centered vortex with $R \leq 1$, 
the vortex is dynamically stable by keeping the small amplitude precession 
around the center. 
With increasing $R$, however, the amplitude of the precession becomes 
larger and larger over time; the vortex spirals out as seen in Fig.~\ref{10vortexstripe}(b). 
Concurrently, the vortex in the other component enters from the outer region. 
This conversion dynamics of the vortex is a consequence of the conservation of the total angular 
momentum, and actually seen in usual two-component BECs \cite{Garcia,Skyrbin}. 
For a centered vortex, a criteria of the dynamic stability is given by 
$g > g_{12}$, which is satisfied in our case. 
The stability of the off-centered vortices has not been discussed so far in two-component BECs, 
but our results indicate that there is some critical displacement from the center 
for a vortex to be dynamically unstable.  
The fluctuation of the trajectory seen in Fig.~\ref{10vortexstripe}(b) is a consequence 
of the density modulation in the stripe phase. 

\begin{figure}[ht]
\centering
\includegraphics[width=1.0\linewidth,bb=0 0 595 396]{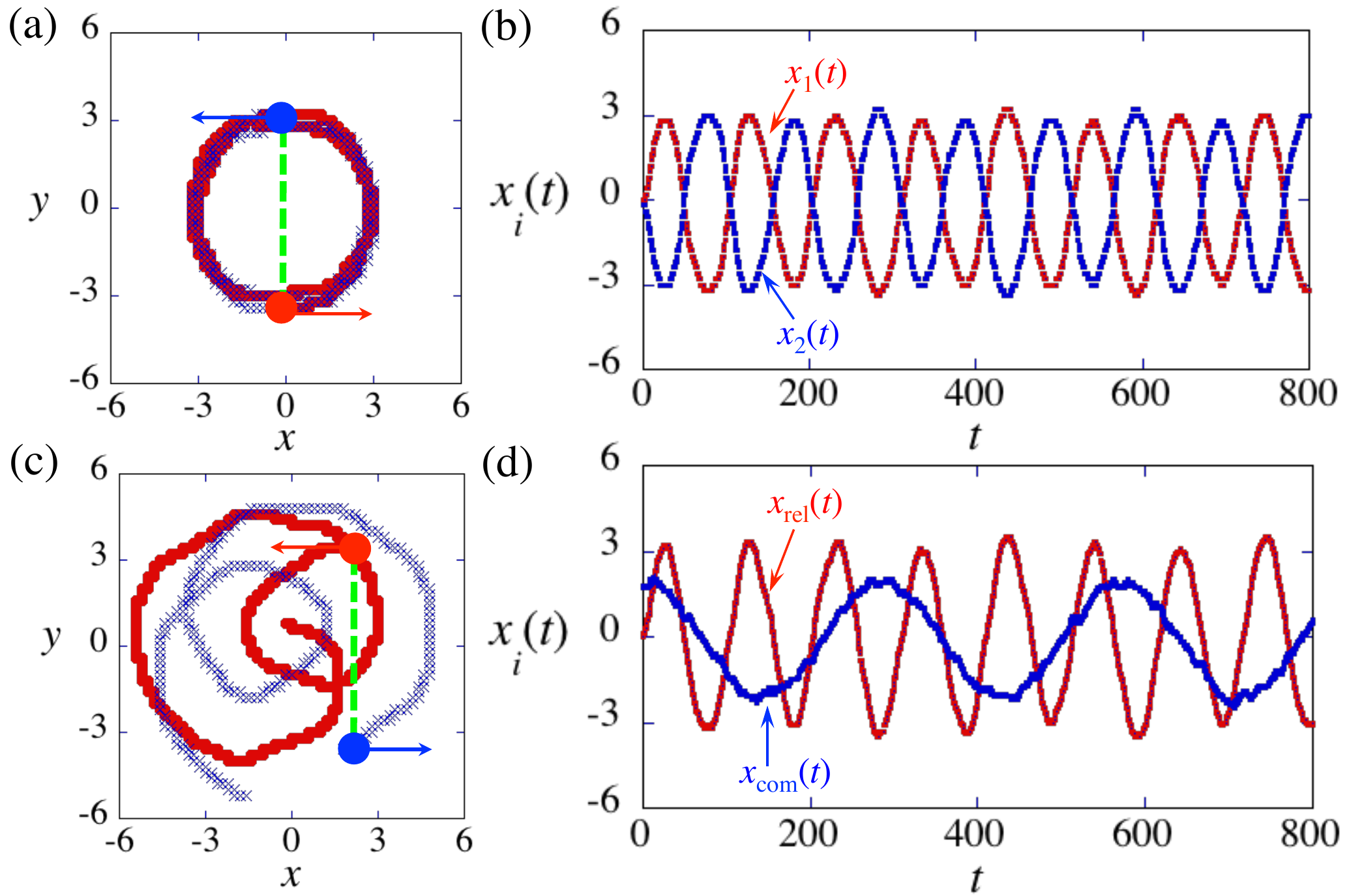} \\\
\caption{(Color online) The vortex motion for $(q_1,q_2) = (1,1)$ and $\Omega=0.2$. 
The vortex molecule is initially polarized along the $y$-direction, in which 
the molecular length is 6.5 (in unit of $k_0^{-1}$). 
(a) The trajectories of each vortex. The center-of-mass 
of the vortex molecule are located on $(x_{\mathrm{com}},y_{\mathrm{com}})=(0,0)$. 
The (red) circles and (blue) crosses represent the vortex positions in 
the $\psi_1$- and $\psi_2$-component, respectively. 
(b) The time development of the $x$-coordinates of the vortices in 
the $\psi_1$- and $\psi_2$-component, corresponding to (a). 
(c) The same as (a) but the initial position of the center-of-mass is 
$(x_{\mathrm{com}},y_{\mathrm{com}})=(2,0)$.
(d) The corresponding time development of the $x$-component of 
the center-of-mass and relative coordinates of the vortex molecule, corresponding to (c). }
\label{11stripevor}
\end{figure}

Next, we turn to the $(q_1,q_2) = (1,1)$ vortex state. In this case, each vortex is bounded 
by a sine-Gordon kink in the relative phase \cite{Son,Kasamatsu,Kasamatsu2,Cipriani}. 
When the center-of-mass of the molecule is 
positioned at the origin, the vortex molecule undergoes self-rotation 
as shown in Fig.~\ref{11stripevor}(a). 
When the center-of-mass is shifted from the origin, the trajectories of each vortex 
show complicated curves [Fig.~\ref{11stripevor}(c)]. However, when we look 
at their center-of-mass coordinate $\mathbf{r}_{\mathrm{com}} = (\mathbf{r}_1 + \mathbf{r}_2)/2$ 
and relative coordinate $\mathbf{r}_{\mathrm{rel}} = (\mathbf{r}_1 - \mathbf{r}_2)/2$, 
the motion can be explained by the simple sinusoidal translation and self-rotational 
motion of the molecule as shown in Fig.~\ref{11stripevor}(d). 

Figure~\ref{11stripevor} clearly shows that the vortex motion consists of two ingredients. 
The center-of-mass motion of the off-centered vortices 
can be caused by the density inhomogeneity due to the trapping potential, where 
the velocity of the motion is proportional to $\hat{\mathbf{z}} \times \nabla V_{\mathrm{tr}}$ 
\cite{Fett09,Fetter}. The analytical form of the precession frequency of a off-centered vortex 
is known as $\Omega = \Omega_m / (1-r_v^2/R_{\perp}^2)$, where 
$\Omega_{m} = (3 \hbar/2mR_{\perp}^2) \ln (R_{\perp}/\xi)$ is the critical rotation 
frequency that gives (meta)stability of a centered vortex and $R_{\perp}$ is a 
Thomas-Fermi radius along the radial direction. 
Rewriting this to the dimensionless form and 
using $r_v =2$, $R_{\perp} \approx 20$ and $\xi \sim 1$, we get a rough estimation 
of the precession period $T \approx 300$ (in unit of $\hbar/E_{R}$), which agrees with 
the numerical results of $x_{\mathrm{com}}$ in Fig.~\ref{11stripevor}. 
On the other hand, the internal rotation of the vortex molecule can be caused by 
a balance of the Magnus force and the intervortex force, which consists of 
a tension of the sine-Gordon kink and the density-density interaction between 
the two components \cite{Pitaevskiitalk}. 
The tension of the sine-Gordon kink is given by $\sigma = \bar{n} \sqrt{8\hbar^3\Omega/m}$ \cite{Son}. 
The intervortex force caused by the intercomponent density-density interaction 
is repulsive because of $g_{12} > 0$ and asymptotically has a dependence 
$\sim (\ln R/\xi - 1/2)/R^3$ \cite{Eto}. 
The net force for each vortex can be written as $F_{v} \sim \sigma R - (\ln R/\xi - 1/2)/R^3$. 
The latter contribution is smaller than the former when $R$ becomes larger, 
so we neglect this in the following estimation. 
Then, the vortex coordinate $\mathbf{r}_i$ for the $i$-th component 
is governed by the equation of motion 
$2 \pi q_i \dot{\mathbf{r}}_i \times \hat{\mathbf{z}} = \mathbf{F}_v$ \cite{Kasapre}. 
Using the formula of the tension, we get $T \approx 100$, which is fairly agreement 
with the numerical result. 


\subsection{Plane-wave phase}
\begin{figure}[ht]
\centering
\includegraphics[width=1.0\linewidth,bb=0 0 311 311]{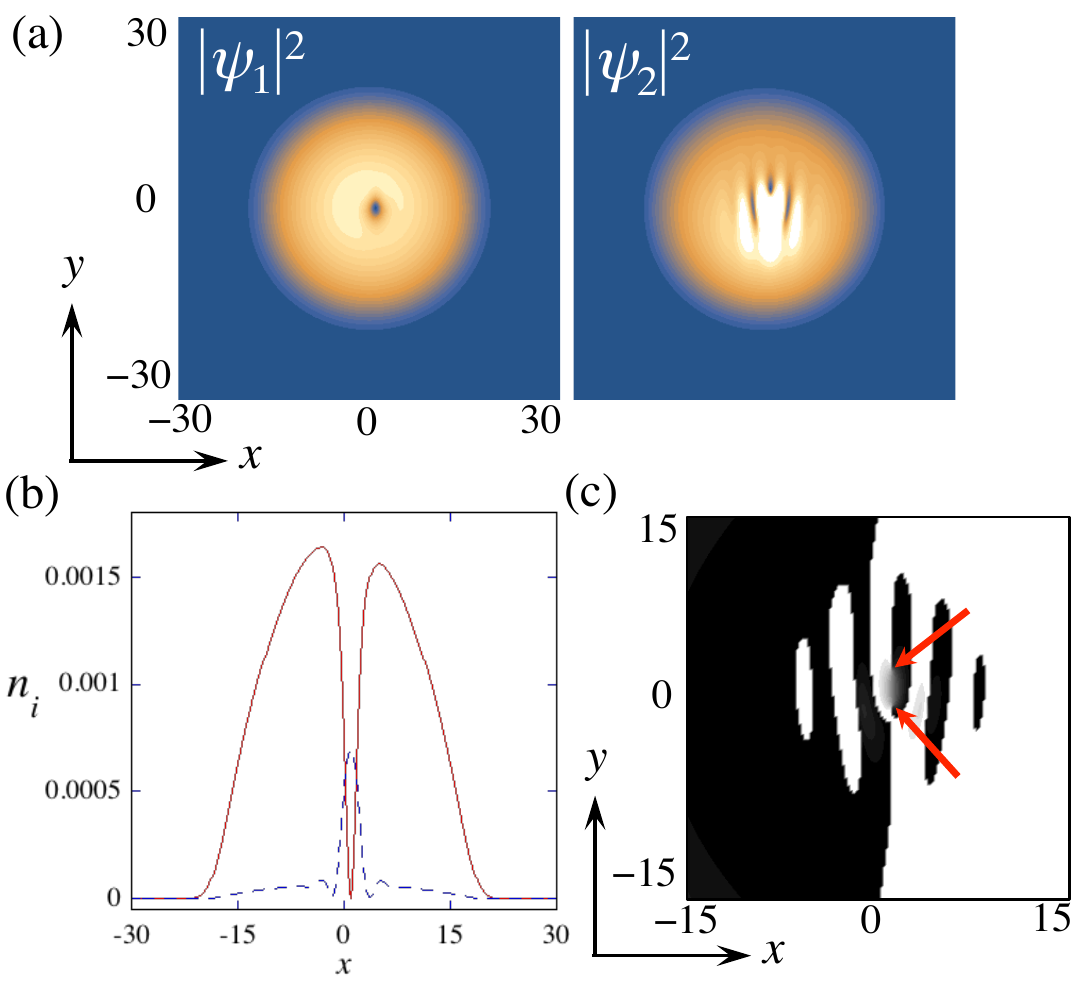} \\\
\caption{Typical initial state used in the simulations in the plane-wave phase. 
Here we choose $(C_{+},C_{-} ) =(0,1)$ ground state with $\Omega = 1.6$ and put 
the $(q_1,q_2) = (1,1)$ vortex at $(x,y)\approx(1,0)$ in the $\psi_1$-component 
and $(x,y)\approx(1,3)$ in the $\psi_2$-component. 
(a) The contour plot of the condensate densities. 
(b) The cross section along $y=0$, where $|\psi_1|^2$ and $|\psi_2|^2$ are plotted 
by solid and dashed curves, respectively. 
(c) The profile of the relative phase. The phase value changes from $-\pi$ (black) 
to $\pi$ (white) continuously. The positions of the vortices are marked by arrows.
}
\label{planeinitial}
\end{figure}

Next, we go to the discussion on the vortex dynamics in the plane-wave phase. 
A quite different point from the stripe phase is that the initial condition for the vortex 
is only limited to $(q_1,q_2) = (1,1)$. 
This is because the relative phase is locked to be $\pi$ over the space because of the 
energy constraint of $E_{\mathrm{Rabi}}$. 
Thus, the individual putting of the vortex on each component is prohibited. 
The typical initial vortex configuration of the simulation is shown in Fig.~\ref{planeinitial}. 
We prepare a slightly off-centered vortex to promote the precession motion 
of the vortex.
The dominant component has a well defined vortex, while the minor component 
has a fragile vortex which accompanies the density modulation around the core. 
The positions of the vortex cores are slightly displaced from each other; the vortex 
core is filled by the density of the other component [Fig.~\ref{planeinitial}(b)]. 
As seen in Fig.~\ref{planeinitial}(c), although the relative phase 
is nearly constant over the space, the vortices are bounded 
by the small sine-Gordon kink. Also, a small modulation of the relative phase 
around the vortex core can be seen in Fig.~\ref{planeinitial}(c).
This may be a remnant of the 
stripe phase, in which the kinetic-energy term with SO coupling likes to modulate the 
wave function. This spatial modulation may be also related with the roton-like 
excitation in the plane-wave phase \cite{Martone}. The density modulation around the 
vortex core is seen in vortices in a dipolar BEC \cite{Pu}, which has also a roton-like minimum 
in the excitation spectrum \cite{Lahaye}. 

\begin{figure}[ht]
\centering
\includegraphics[width=1.0\linewidth,bb=0 0 737 524]{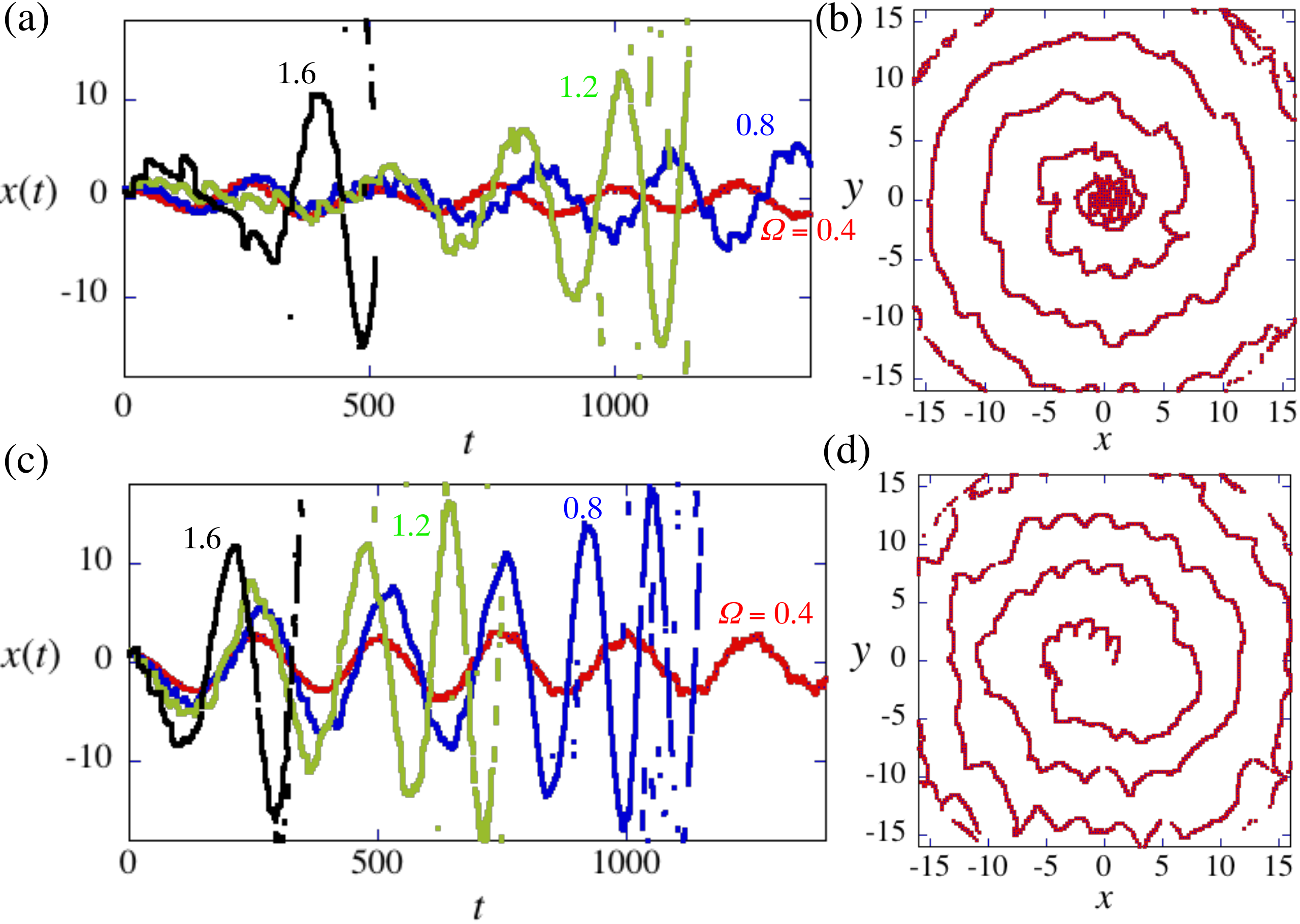} \\\
\caption{The time development of the vortex position in the plane-wave phase. 
The initial position of the vortex in the dominant component is $(x,y) \approx (1,0)$  
The vortex trajectories in the minor component are not shown because the motion 
exhibits turbulent-like behavior so that it is difficult to identify the individual 
vortex trajectory. (a) Time development of $x_v(t)$ for $(C_{+},C_{-}) = (0,1)$ 
with $\Omega=0.4$, 0.8, 1.2, and 1.6. 
(b) The vortex trajectory corresponding to (a) for $\Omega=1.2$. 
(c) Time development of $x_v(t)$ for $(C_{+},C_{-}) = (1,0)$ 
with $\Omega=0.4$, 0.8, 1.2, and 1.6.
(d) The vortex trajectory corresponding to (c) for $\Omega=1.2$. 
}
\label{planedyn}
\end{figure}
\begin{figure}[ht]
\centering
\includegraphics[width=1.0\linewidth,bb=0 0 720 226]{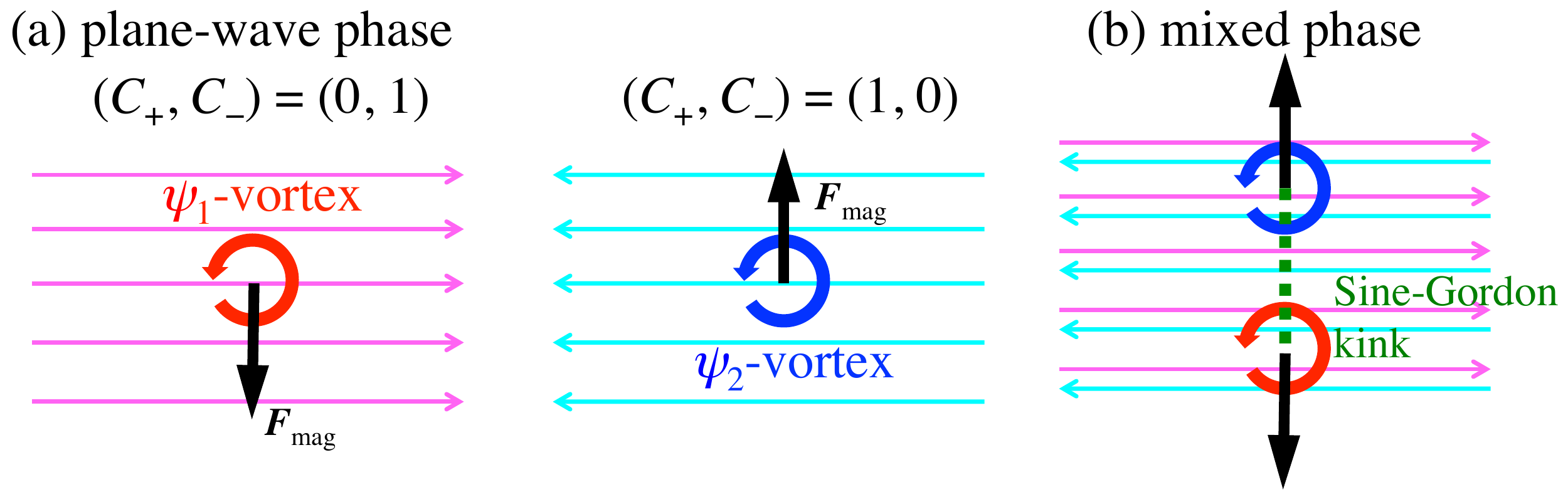} \\\
\caption{Schematic illustration of the vortex propertes in (a) the plane-wave phase 
and (b) the mixed phase. In the plane-wave phase, we only show the dominant component. 
For $(C_+,C_{-}) = (1,0)$, there is a background velocity along $+x$-direction, 
thus the Magnus force for a vortex with $q=1$ directs to $-y$. 
For $(C_+,C_{-}) = (0,1)$, there is a background velocity along $-x$-direction, 
thus the Magnus force for a vortex with $q=1$ directs to $+y$.
In the mixed phase (b), both components are equally populated 
and each vortex has the Magnus force with the same magnitude 
but the opposite direction. The vortex configuration is kept by the balance 
between the Magnus force and the tension of the sine-Gordon kink. 
Thus, the vortex molecule is polarized along the $y$-direction. 
}
\label{planesche}
\end{figure}

The dynamics of this vortex state is shown in Fig.~\ref{planedyn}. 
We can see mainly two features. 
First, the vortex trajectories in the two degenerate ground states are different 
even when the vortex motion starts from the same initial positions. 
The upper panel in Fig.~\ref{planedyn} corresponds to the vortex dynamics for 
$(C_+,C_-) = (0,1)$ (condensation onto  $-\mathbf{k}_1$ momentum state), 
while the lower one does $(C_+,C_-) = (1,0)$ (condensation onto $+\mathbf{k}_1$ momentum state). 
The amplitude of the precession in the upper panel looks to be smaller than that in the lower one.  
The asymmetry of the vortex dynamics can be understood from the 
kinetic energy Eq.~(\ref{kinesocene}). 
For $(C_{+},C_{-}) = (0,1)$, the energy can be written as 
\begin{align}
E_{k-} =  \int d^2 r n(r) \biggl\{ \cos^2 \theta \left[ (v_{1x}+1-k_1)^2 + v_{1y}^2  \right] \nonumber \\ 
+  \sin^2 \theta \left[ (v_{2x}-1-k_1)^2 + v_{2y}^2  \right] \biggr\} ,
\end{align}
and for $(C_{+},C_{-}) = (1,0)$ 
\begin{align}
E_{k+} = \int d^2 r n(r) \biggl\{ \sin^2 \theta \left[ (v_{1x}+1+k_1)^2 + v_{1y}^2 \right] \nonumber \\
+ \cos^2 \theta \left[ (v_{2x}-1+k_1)^2 + v_{2y}^2 \right] \biggr\} ,
\end{align}
where the term proportional to $\cos^{2} \theta$ corresponds to the contribution from the dominant 
component. Then, we can consider that a vortex in the dominant component 
is dragged by a constant background velocity $\mathbf{v}_\mathrm{bg}=(1-k_1) \hat{\mathbf{x}}$ 
for $(C_{+},C_{-}) = (0,1)$
and $\mathbf{v}_\mathrm{bg}=-(1-k_1) \hat{\mathbf{x}}$ for $(C_{+},C_{-}) = (1,0)$ 
because of the Magnus effect. This situation is schematically shown in Fig.~\ref{planesche}(a). 
For $(C_+,C_-) = (0,1)$ $\mathbf{v}_\mathrm{bg}$ is opposite to the 
counterclockwise direction of the precession motion caused by the density gradient, 
so that the precession motion tends to be suppressed. 
On the other hand, for $(C_+,C_-) = (1,0)$ 
$\mathbf{v}_\mathrm{bg}$ enhances the tendency of the counterclockwise precession. 

Secondly, we can see the significant damping behavior of the vortex motion, 
which is typical for the vortex precession in the presence of dissipation \cite{Fett09}. 
The vortex spirals out to outward as seen in Fig.~\ref{planedyn}(b) and (d). 
Different from the motion seen in the stripe phase [Fig.~\ref{10vortexstripe}], 
the vortex is not converted to the other component. 
However, this is questionable because dissipation is absent in our problem. 
One reason of this damping behavior may be due to the presence 
of the minor component. Because the amplitude of the minor component is 
very small, the wave function is highly excited to undergo a turbulent-like state 
during the time evolution, which plays the role of thermal noise for the other dominant component. 
Actually, the vortex decays faster and faster when the Rabi coupling 
is increased for the minor component to be more populated [see Fig.~\ref{planedyn}(a) and (c)]. 
However, as the system approaches to the mixed phase with increasing $\Omega$ further, 
the balance of the Magnus force and the tension of the sine-Gordon kink prohibits 
the vortices to go outward, effective dissipation mechanism thus being no more present 
in the mixed phase as shown below. 

\subsection{Mixed phase}
\begin{figure}[ht]
\centering
\includegraphics[width=1.0\linewidth,bb=0 0 680 226]{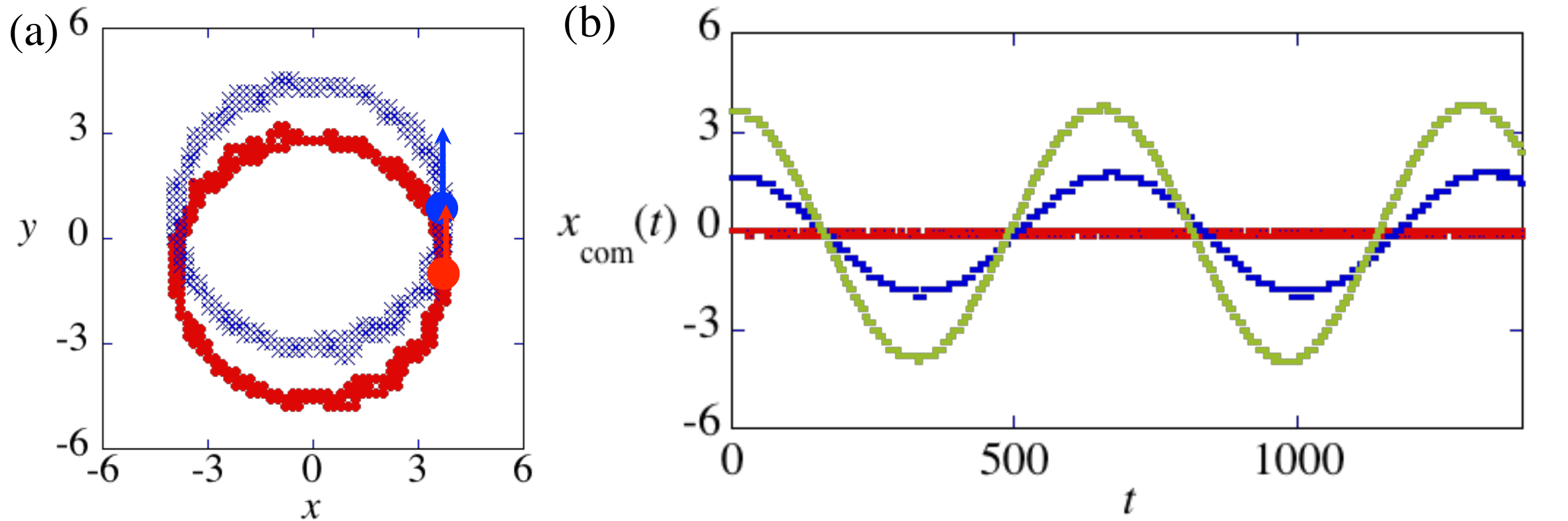} \\\
\caption{The time development of the vortex position in the mixed phase 
with $\Omega = 4.4$. 
(a) The vortex trajectories of $\psi_1$- (red circles) and $\psi_2$-component (blue crosses). 
The initial position is indicated by large circles. 
(b) The time development of the $x$-coordinate of the center-of-mass of the two vortices, where 
the initial positions of the center-of-mass are changed as $x_{\mathrm{com}} = 0$, 1.75, 3.5.  
}
\label{mixedvor}
\end{figure}

Finally, we will discuss the vortex dynamics in the mixed phase. 
Since there is a strong Rabi coupling in this case, 
the $(q_1,q_2) = (1,1)$ vortices are strongly bounded 
to form the vortex molecule. 
However, different from that seen in the stripe phase [Fig.~\ref{11stripevor}], 
the vortex molecule does not show the internal rotation; 
its polarization is fixed to be the $y$-direction. 
In this phase, the energy is given by
\begin{align}
E_{k0} =  \int d^2 r n(r) \biggl\{\frac{1}{2} \left[ (v_{1x}+1)^2 + v_{1y}^2  \right] \nonumber \\ 
+  \frac{1}{2} \left[ (v_{2x}-1)^2 + v_{2y}^2  \right] \biggr\} .
\end{align}
Thus, the Magnus force caused the background velocity induced by the SO coupling 
has the same magnitude and the opposite direction. Since the vortices are attracted by 
the tension of the sine-Gordon kink, the exact balance between 
Magnus force and the kink tension stabilizes dynamically the vortex 
molecule at the same position.   
For the off-centered vortex molecule, the center-of-mass of the vortex molecule 
shows a circular motion around the origin without rotating the molecule itself, 
as shown in Fig.~\ref{mixedvor}. 

\begin{figure}[ht]
\centering
\includegraphics[width=1.0\linewidth,bb=0 0 538 368]{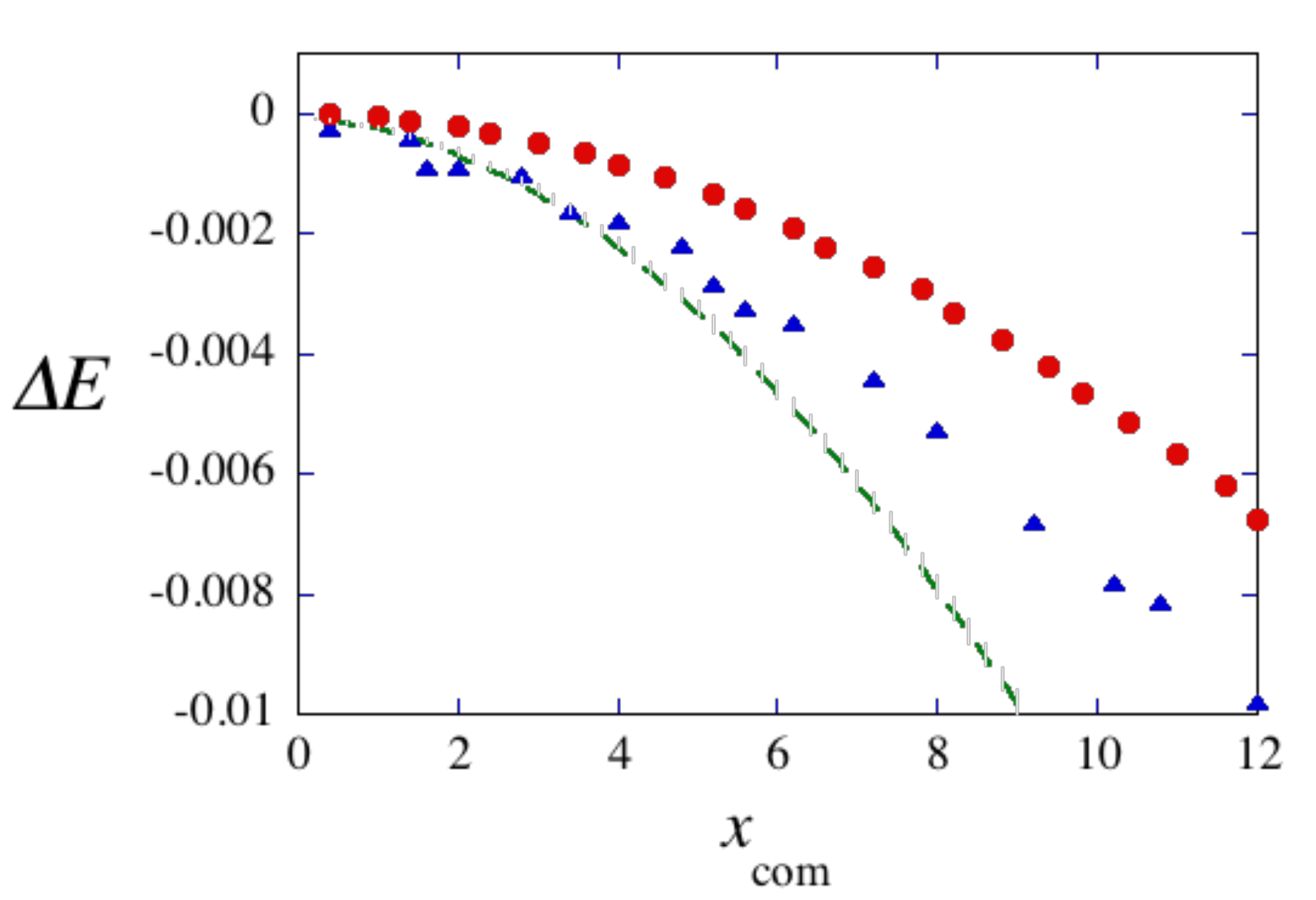} \\\
\caption{The total energy as a function of the center-of-mass displacement of the 
vortex molecule from the center. $\Delta  E$ is defined by 
$\Delta E =  E(x_{\mathrm{com}}) - E(0) $. 
The energy is calculated from the states with an off-center vortex molecule, 
obtained by a similar way as described in the 1st paragraph in Sec.~\ref{nume}. 
The circles represent the energy of the off-centered vortex molecule 
in the mixed phase with $\Omega=4.4$, while the triangles for that 
in the striped phase with $\Omega=0.2$. 
The dashed curve shows the 
energy of an off-centered vortex in a single component BEC, where we consider 
the situation $N=N_1$ ($N_2=0$) without the SO coupling. 
}
\label{vorenergy}
\end{figure}
Note that the period of the center-of-mass is longer than that observed 
in the stripe phase in Fig.~\ref{11stripevor}. 
The period of the vortex precession is basically determined by the gradient of 
the total energy with respect to the vortex position $\mathbf{r}_v$ \cite{Fett09,Mcgee}. 
Here, we plot the energy as a function of the center-of-mass position of the vortex 
molecule in Fig.~\ref{vorenergy}. 
For comparison, we also plot the same energy in the stripe phase ($\Omega = 0.2$) 
as well as for an off-center vortex in a single-component BEC. 
The energy for the stripe phase is more or less agreed with that for a single-component 
BEC, especially for $x_{\mathrm{com}} < 5$. This is the reason why the precession period can be estimated 
by the known formula as discussed in Sec.~\ref{stripevortexdy}. 
On the other hand, the slope of the energy in the mixed phase decays slower than the other cases. 
This indicates that the vortex molecule are stabilized against the center-of-mass 
displacement because of the balance of the forces shown in Fig.~\ref{planesche}(b). 

\section{discussion and conclusion}\label{concle}
We study the vortex dynamics in laser-induced SO coupled BECs. 
The characteristics of the vortex dynamics is very different in each ground state phase, 
which can be used to identify each phase. 
The vortex dynamics in the stripe phase is similar to those in the conventional 
two-component BECs, while the plane-wave phase has asymmetric vortex dynamics 
in the two degenerate ground state. 
In the mixed phase, the polarization of the vortex molecule is kept during the 
dynamics, which is very different behavior 
seen in the conventional two-component BEC with only coherent Rabi coupling. 
We hope that these numerical solutions would be useful for better understanding 
of the vortex dynamics in exotic superfluids with SO coupling. 
However, we would like to point out that the understanding of the vortex dynamics in 
conventional two-component BECs is still incomplete, although there has 
been some progress recently \cite{Pitaevskiitalk,Kasapre,Etotalk}. 

\section*{Acknowledgement} 
The author thanks Luis Santos and all members in his group 
for hospitality in Leibniz University Hannover. 
This work was partly supported by KAKENHI from JSPS (Grant No. 26400371).


\begin{thebibliography}{99}

\bibitem{Lin11}  
Y.-J. Lin,  K. Jim{\'e}nez-Garc{\'\i}a, and I. B. Spielman, 
Nature(London) \textbf{471}, 83 (2011). 

\bibitem{Galitskirev}
V. Galitski	and I. B. Spielman, 
Nature(London) \textbf{494}, 49 (2013). 

\bibitem{Goldmanrev}
N. Goldman, G. Juzeliunas, P. Ohberg, I. B. Spielman
Rep. Prog. Phys. \textbf{77}, 126401 (2014). 

\bibitem{Zhairev}
H. Zhai, 
Rep. Prog. Phys. \textbf{78}, 026001 (2015). 

\bibitem{Fett09}  
A. L. Fetter, 
Rev. Mod. Phys. \textbf{81}, 647 (2009). 

\bibitem{Wu}
C. Wu and I. Mondtragon-Shem, 
Chin. Phys. Lett. \textbf{28}, 097102 (2011). 

\bibitem{Sinha}
S. Sinha, R. Nath, and L. Santos, 
Phys. Rev. Lett. \textbf{107}, 270401 (2011). 

\bibitem{Hui}
H. Hu, B. Ramachandhran, H. Pu, and X.-J. Liu, 
Phys. Rev. Lett. \textbf{108}, 010402 (2012). 

\bibitem{Radi11}  
J.\ Radi{\'c}, T. A. Sedrakyan, I.B. Spielman, and V. Galitski, 
Phys. Rev. A {\bf 84}, 063604 (2011). 

\bibitem{Xu}
X.-Q. Xu and J. H. Han, 
Phys. Rev. Lett. \textbf{107}, 200401 (2011). 

\bibitem{Zhou}
X.-F. Zhou, J. Zhou, and C. Wu, 
Phys. Rev. A \textbf{84}, 063624 (2011). 

\bibitem{Zhang}
J.-Y. Zhang, S.-C. Ji, Z. Chen, L. Zhang, Z.-D. Du, B. Yan, G.-S. Pan, B. Zhao, Y.-J. Deng, H. Zhai, S. Chen, and J.-W. Pan, 
Phys. Rev. Lett. \textbf{109}, 115301 (2012).

\bibitem{Wang}
P. Wang, Z.-Q. Yu, Z. Fu, J. Miao, L. Huang, S. Chai, H. Zhai, and J. Zhang, 
Phys. Rev. Lett. \textbf{109}, 095301 (2012).

\bibitem{Cheuk}
L. W. Cheuk, A. T. Sommer, Z. Hadzibabic, T. Yefsah, W. S. Bakr, and M. W. Zwierlein, 
Phys. Rev. Lett. \textbf{109}, 095302 (2012).

\bibitem{Huang}
L. Huang, Z. Meng, P. Wang, P. Peng, S.-L. Zhang, L. Chen, D. Li, Q. Zhou, J. Zhang, 
arXiv:1506.02861 (2015).

\bibitem{Ho}
T.-L. Ho and S. Zhang, 
Phys. Rev. Lett. \textbf{107}, 150403 (2011). 

\bibitem{Li}
Y. Li, L. Pitaevskii and S. Stringari, 
Phys. Rev. Lett. \textbf{108}, 225301 (2012). 

\bibitem{Martone}
G. I. Martone, Y. Li, L. Pitaevskii and S. Stringari, 
Phys. Rev. A \textbf{86}, 063621 (2012). 

\bibitem{Lu13}   
Q.-Q. L\"{u} and D. E. Sheehy, 
Phys. Rev. A \textbf{88}, 043645 (2013).

\bibitem{Li2}
Y. Li, G. I. Martone, L. Pitaevskii and S. Stringari, 
Phys. Rev. Lett. \textbf{110}, 235302 (2013). 

\bibitem{Ozawa}
T. Ozawa, L. P. Pitaevskii, and S. Stringari, 
Phys. Rev. A \textbf{87}, 063610 (2013). 


\bibitem{Hamner}
C. Hamner, Y. Zhang, M. A. Khamehchi, M. J. Davis, and P. Engels,
Phys. Rev. Lett. \textbf{114}, 070401 (2015).

\bibitem{Fetter} 
A. Fetter, 
Phys. Rev. A \textbf{89}, 023629 (2014). 

\bibitem{Weiler}
C.N. Weiler, T.W. Neely, D.R. Scherer, A.S. Bradley, M.J. Davis, B.P. Anderson, 
Nature (London) {\bf 455}, 948 (2008).

\bibitem{Frei10}   
D.V. Freilich, D.M. Bianchi, A.M. Kaufman, T.K. Langin, and D.S. Hall, 
Science {\bf 329}, 1182 (2010). 

\bibitem{tyyu}
The value of $N$ represents the particle number in the 2D plane, where 
the condensate is assumed to be homogeneous along the $z$-axis, and 
can be written as $N=N_{\mathrm{3D}}/L_z$ with the typical size $L_z$ 
of the $z$-direction.

\bibitem{Abad}
M. Abad and A. Recati, 
Euro. Phys. J. D \textbf{67}, 148 (2013). 

\bibitem{chu2}
We would like to point out that the variational wave function introduced by Fetter \cite{Fetter} 
to study the vortex dynamics in the Lagrangian formulation is applicable only to the plane-wave phase. 
His trial function for the spinor is given by 
\begin{equation}\label{zeta}
\zeta = \left(
\begin{matrix}  e^{i \theta_1} \cos \theta\\
- e^{i \theta_2} \sin \theta
\end{matrix}\right)  e^{i \alpha x}
\end{equation}
where $\theta_j = m_j \arctan\left(\frac{y-y_0}{x-x_0}\right)$ represents the vortex winding.
An additional velocity contribution $\alpha x$ was introduced to account for the spatial asymmetry of $h_0$. 
This ansatz is nothing but the wave function of the plane-wave phase of Eq.~(\ref{planewavewav}), 
which is applicable for the range $0.3 \leq \Omega \leq 4$. For small $\Omega (\leq 0.3)$, 
where the stripe phase is the ground state, the ansatz Eq.~(\ref{zeta}) cannot describe 
the dynamics correctly. 

\bibitem{chu1}
The vortex state $(q_1,q_2)=(1,0)$ in the stripe phase is not energetically 
preferable for the condensates with a intercomponent coherent coupling. 
During the imaginary time evolution, the interconversion of the particle number can 
make the non-vortex component more populated, which reduces the total energy, 
and eventually eliminates the component with a vortex. 
Thus, we fix each particle number during the second imaginary time propagation 
in this case. 

\bibitem{Son}
D. T. Son and M. A. Stephanov, 
Phys. Rev. A \textbf{65}, 063621 (2002).

\bibitem{Kasamatsu}
K. Kasamatsu, M. Tsubota, and M. Ueda, 
Phys. Rev. Lett. \textbf{93}, 250406 (2004).

\bibitem{Kasamatsu2}
K. Kasamatsu, M. Tsubota, and M. Ueda
Phys. Rev. A \textbf{71}, 043611 (2005). 

\bibitem{Cipriani}
M. Cipriani and M. Nitta
Phys. Rev. Lett. \textbf{111}, 170401 (2013).

\bibitem{Garcia}
V. M. P\'{e}rez-Garc\'{i}a and J. J. Garc\'{i}a-Ripoll, 
Phys. Rev. A \textbf{62}, 033601 (2000).

\bibitem{Skyrbin} 
D. V. Skryabin
Phys. Rev. A \textbf{63}, 013602 (2000).

\bibitem{Pitaevskiitalk}
L. Pitaevskii, talk at ``Cold Atoms Meet High Energy Physics", 2015, Trento Italy.

\bibitem{Eto}
M. Eto, K. Kasamatsu, M. Nitta, H. Takeuchi, M. Tsubota, 
Phys. Rev. A \textbf{83}, 063603 (2011).

\bibitem{Kasapre}
K. Kasamatsu, M. Eto, and M. Nitta, 
in preparation. 


\bibitem{Pu}
S. Yi and H. Pu, 
Phys. Rev. A \textbf{73}, 061602(R) (2006).

\bibitem{Lahaye}
T. Lahaye, C. Menotti, L. Santos, M. Lewenstein, T. Pfau,
Rep. Prog. Phys. \textbf{72}, 126401 (2009). 

\bibitem{Mcgee}
S. A. McGee and M. J. Holland
Phys. Rev. A \textbf{63}, 043608 (2001).

\bibitem{Etotalk}
M. Eto and M. Nitta, talk at ``Cold Atoms Meet High Energy Physics", 2015, Trento Italy.

\end{thebibliography}
\end{document}